\documentclass[prb,amsmath,amssymb,twocolumn,floatfix]{revtex4}

\usepackage{graphicx}
\usepackage{subfigure}
\usepackage{dcolumn}
\usepackage{verbatim}
\usepackage{epstopdf}

\def\dd{\textrm{d}}
\def\Tr{\textrm}
\def\Bf{\boldsymbol}

\def\rv{\Bf{r}}

\def\Pv{\Bf{P}}

\def\Av{\Bf{A}}

\def\nv{\Bf{\nabla}}
\def\xh{\hat{\Bf{x}}}

\begin{document}
\title{Two-dimensional fermionic superfluids, pairing instability and vortex liquids in the absence of time reversal symmetry}

\author{Predrag Nikoli\'c}
\affiliation{Department of Physics, Rice University, Houston, TX 77005}

\begin{abstract}

We consider a generic two-dimensional system of fermionic particles with attractive interactions and no disorder. If time-reversal symmetry is absent, it is possible to obtain incompressible insulating states in addition to the superfluid at zero temperature. The superfluid-insulator phase transition is found to be second order in type-II systems using a perturbative analysis of Cooper pairing instability in quantum Hall states of unpaired fermions. We obtain the pairing phase diagram as a function of chemical potential (density) and temperature. However, a more careful analysis presented here reveals that the pairing quantum phase transition is \emph{always} preempted by another transition into a strongly correlated normal state which retains Cooper pairing and cannot be smoothly connected to the quantum Hall state of unpaired fermions. Such a normal phase can be qualitatively viewed as a liquid of vortices, although it may acquire conventional broken symmetries. Even if it did not survive at finite temperatures its influence would be felt through strong quantum fluctuations below a crossover temperature scale. These conclusions directly apply to fermionic ultra-cold atom systems near unitarity, but are likely relevant for the properties of other strongly correlated superfluids as well, including high temperature superconductors.

\end{abstract}

\date{\today}

\maketitle

Many years of efforts to understand the strongly correlated phases of cuprate superconductors and other unconventional materials have resulted in a variety of new conceptual ideas. One broadly accepted idea is that the destruction of two-dimensional superconductivity in underdoped cuprates involves strong fluctuations of the order parameter instead of Cooper pair dissociation \cite{Emery}. The remarkable theoretical insight gained from duality transformations in several systems \cite{Dual1, Dual2} stimulates viewing this phase transition as a proliferation of vortices. The normal state may be envisioned as a liquid of mobile vortex loops. Indeed, there are several experimental observations that support this view \cite{Ong1, Ong2}, and several theories that exploit it \cite{Zlatko1, Subir1} to successfully explain certain unconventional properties of cuprates \cite{Fischer, Yazdani, Hudson, Davis, Valla}.

Nevertheless, the notion of a vortex liquid is not a trivial one. It is not even clear yet whether any correlated state of any known material truly deserves to be called a vortex liquid. In particular, the vortex liquid regime found in the pseudogap region in cuprates seems to be separated from the high temperature normal state only by a crossover. A quantum vortex liquid might be a genuine phase at zero temperature which influences the finite temperature dynamics \cite{QED3, VorLiq1}. A few theoretical models can reliably describe strongly correlated phases that can be characterized as vortex liquids, and they provide the best route to precisely define and characterize these phases \cite{VorLiq1, VorLiq2, VorLiq3}.

A new platform for addressing these interesting challenges in many-body physics has been recently introduced in atomic physics \cite{Bloch}. The ability to condense ultra-cold fermionic atoms into a superfluid with controllable strength of interactions and study them as virtually ideal and clean systems has already mobilized an effort to simulate condensed matter systems. Static three-dimensional cold-atom systems near a broad Feshbach resonance have striking universal properties \cite{Bloch, Kohler, unitary, rvs} which make them amenable to field theoretical analysis. This universality is controlled by a quantum critical point at the zero density Feshbach resonance \cite{unitary}, which defines the unitarity limit. The superfluid in this regime interpolates between the Bardeen-Cooper-Schrieffer (BCS) state of Cooper pairs and Bose-Einstein condensate (BEC) of diatomic molecules, and insulating states in optical lattices interpolate between the two analogous limits of band and Mott insulators respectively \cite{optlat}. The cold-atom superfluids can be destabilized by fast rotation in the same way the electronic superconductor can be destabilized by magnetic field \cite{Moller, ZhaiHo, YangZhai}. Experimental progress toward observing this effect in cold atoms is still limitted, but the overall pace of development in the field gives many reasons for optimism \cite{Dalibard, Engels, Stock}.

\begin{figure}
\subfigure[{}]{\includegraphics[height=1.6in]{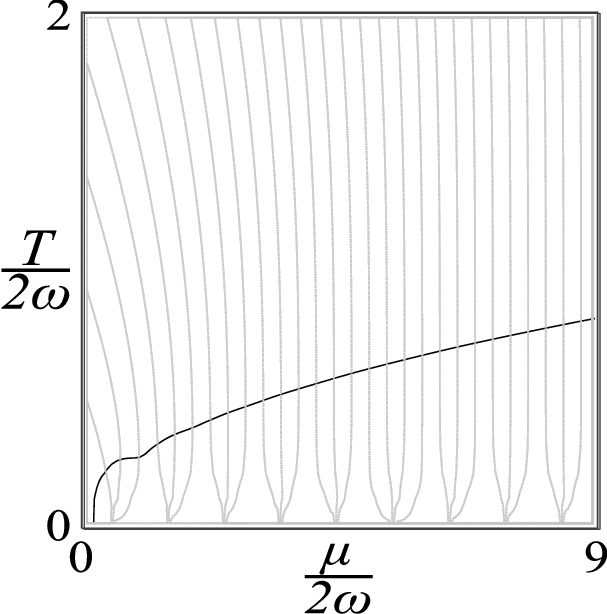}}
\subfigure[{}]{\includegraphics[height=1.6in]{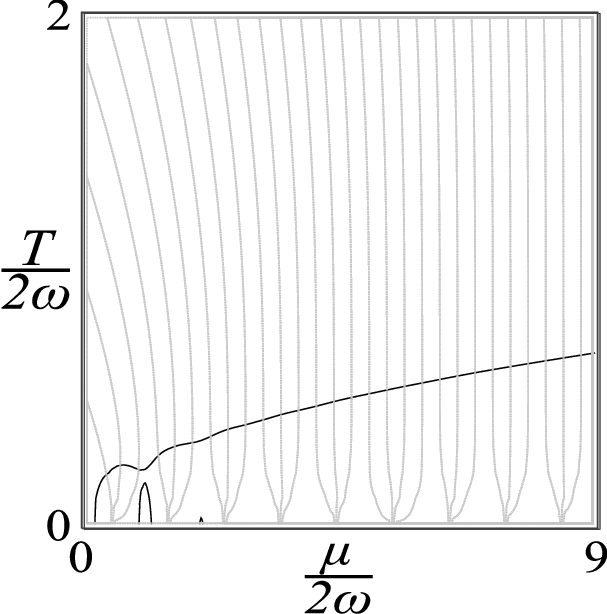}}
\subfigure[{}]{\includegraphics[height=1.6in]{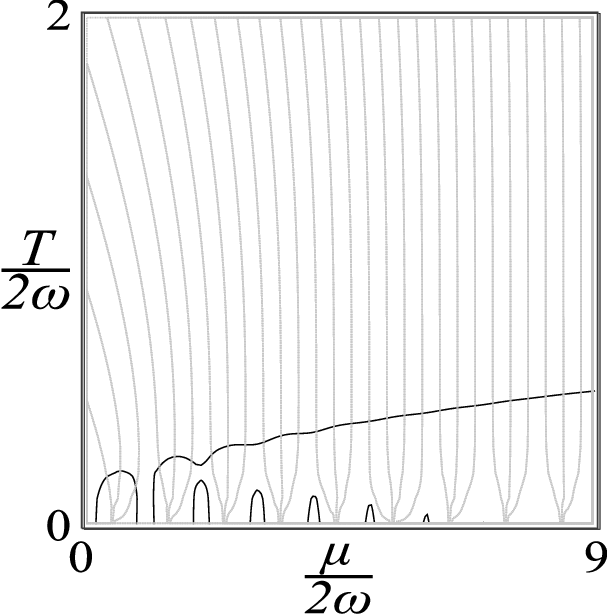}}
\subfigure[{}]{\includegraphics[height=1.6in]{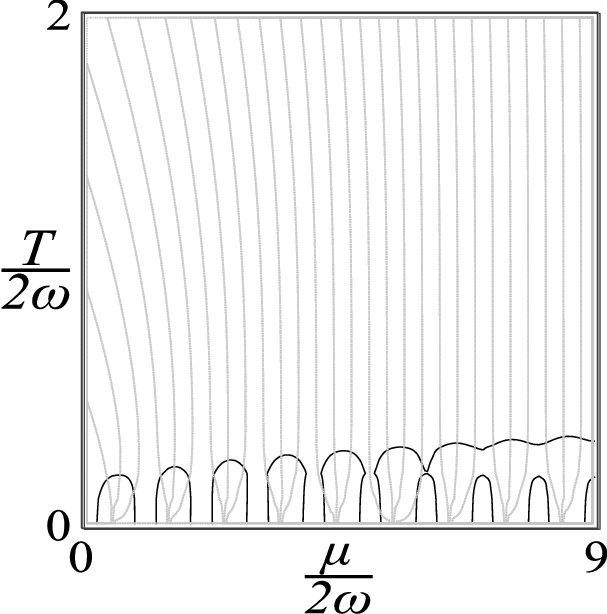}}
\caption{\label{PairingPD}Critical temperature of the superfluid-normal pairing transition as a function of chemical potential. Interactions between particles are characterized by the two-body scattering length $a$ and become weaker going from (a) to (d): $\widetilde{\nu} \propto -a^{-1}$ (defined in the text and Fig.\ref{Pairing2}) is $0.3$ in (a), $0.4$ in (b), $0.5$ in (c), $0.6$ in (d). The vertically stretching lines are normal state constant density contours, with increment $0.25 B/(2\pi)$ where $B=2m\omega$ is the uniform flux density.}
\end{figure}

In this paper we analyze a simple two-dimensional model of interacting fermionic particles which is a significant simplification of electronic systems such as cuprates in magnetic field, but can be routinely and accurately realized using rotating ultra-cold atoms. The model contains paired superfluids at low temperatures and a normal phase of uncorrelated fermions at high temperatures. We concentrate on type-II superfluids which host an Abrikosov vortex lattice due to the absence of time-reversal symmetry. The uncorrelated normal phase is a thermally excited quantum Hall state of unpaired fermions. A characteristic phase diagram of pairing instability in this normal phase is shown in Fig.\ref{PairingPD}. One of our main results is that the normal phase can stretch all the way to zero temperature at finite particle densities for certain strengths of interactions between fermions.

The pairing instability depicted in Fig.\ref{PairingPD} was examined in the past by several authors; for a review see \cite{Zlatko3} and references therein. While three-dimensional systems attracted most attention in the literature, the focus here is on two-dimensions precisely because of the previously missed zero-temperature normal states (they are not expected to occur in three dimensions). In this paper we provide a fresh approach to issues such as superconductivity in high magnetic fields and keep the formalism simple and clear in order to focus on the most fundamental questions. This simplicity will allow us to reach some fundamental conclusions which were obscured in the past by the inability to go beyond the mean-field and other approximations. We will not consider Zeeman effect in this paper; Zeeman pair-breaking effects can be completely eliminated in cold-atom systems, while in electronic systems they are certainly important although the superconducting state may be able to survive them even in high magnetic fields (this analysis will be published separately). Another popular approach in the literature has been the Landau-Ginzburg theory, but for our purposes it is important to take into account the internal structure of the bosonic degrees of freedom responsible for superfluidity and capture phenomena characteristic for fermionic superfluids that are absent in pure bosonic systems and Landau-Ginzburg theories.

\begin{figure}
\includegraphics[height=2.3in]{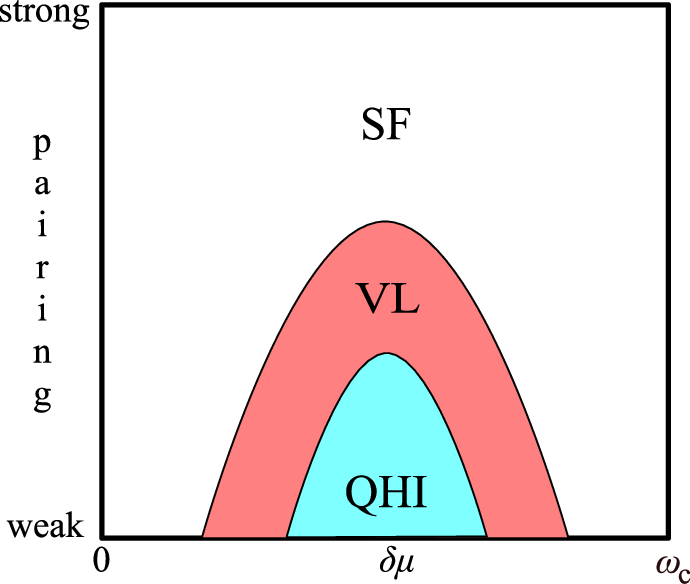}
\caption{\label{VLphdiag}(color online) A schematic phase diagram at zero temperature. The depicted phases are fermionic quantum Hall insulator (QHI), vortex liquid (VL) and superfluid with an Abrikosov vortex lattice (SF). The vertical axis is a measure of the attractive interaction strength between fermions. The horizontal axis in is fermion chemical potential expressed in reference to a Landau level ($\omega_c$ is cyclotron frequency; the left and right edge of the diagram correspond to two neighboring Landau levels). In the weak pairing limit the superfluid can be destabilized between Landau levels, forming a vortex liquid across a first order phase transition. For very weak pairing interactions a quantum Hall insulator of unpaired fermions might become stable across a transition such as the one discussed in Ref.\cite{YangZhai}. The particle densities in VL and QHI are expected to be fixed and commensurate with respect to magnetic area. In this paper we find integer QHI states shown in Fig.\ref{PairingPD} and the corresponding VL states around them. Other insulating states such as fractional bosonic and fermionic quantum Hall states can be expected in various circumstances.}
\end{figure}

The possibility of finding a zero-temperature normal phase of unpaired fermions, which is actually a quantum Hall insulator, provides an opportunity to establish with physical rigor the existence of another non-superfluid phase at zero temperature. We show that the zero-temperature superfluid at a finite density must be destroyed by the quantum fluctuations of the order parameter before the depairing instability occurs. This vortex lattice melting quantum phase transition is expected to be first order \cite{Nelson, Herbut, Franz}, although proposals for a second order transition have also been made \cite{Radzihovsky}. The resulting normal phase is strongly correlated as it retains Cooper pairs, and it cannot be smoothly connected with the quantum Hall state of unpaired fermions. Quantum phase transitions between the two kinds of normal phases have been studied recently \cite{YangZhai}. We will refer to the new normal phase as vortex liquid, based on its sharp distinction from the fermionic quantum Hall states. A schematic phase diagram is shown in Fig.\ref{VLphdiag}.

The concrete properties of a vortex liquid and its stability at finite temperatures are not addressed in this paper. The main reason is that these properties are not universal in generic realistic systems. To make this point we use a renormalization group analysis and show that all possible interactions are relevant at the Gaussian fixed point in two dimensions, regardless of their spatial dependence or the number of colliding particles. The existence of infinitely many relevant directions in the parameter space suggests the possible existence of a large number of stable interacting fixed points which determine the properties of phases. Which of these phases is realized in a particular realistic system is highly sensitive to the microscopic details of interactions between particles. One prominent candidate in the case of short-range interactions is a quantum Hall liquid of incoherent Cooper pairs \cite{YangZhai}. Other possibilities include a conventional broken symmetry such as a density wave, in which case its algebraically correlated analogue could exist as a stable phase at finite temperatures.

Since the ultra-cold fermionic atoms near unitarity are the closest realization of the model considered here, we will devote extra attention to their specifics and express the results in scales natural to cold atoms. However, the analysis is general and possibly relevant to other strongly correlated systems such as cuprates.

This paper is organized into several sections. We begin by defining the model of interest in section \ref{SecModel} and setting up the formalism for perturbation theory. In section \ref{SecPairing} we analyze the phase diagram of the Cooper pairing transition in the perturbation theory. The interpretation and limitations of the perturbative results are outlined in section \ref{SecLim}. In section \ref{SecVL} we establish that vortex liquid phases, missed by the naive perturbation theory, exist at zero temperature. At the end, we apply a renormalization group analysis in section \ref{SecRG} to appreciate the non-universal properties of the vortex liquid states and the richness of its possible physical realizations. We summarize the results and discuss a possible experimental approach in the concluding section \ref{SecCon}.

\subsection{The model}\label{SecModel}

Our starting point is a generic model of fermionic particles with attractive interactions in two dimensions. We couple the particles to a U(1) gauge field which implements a uniform flux density $B$:
\begin{equation}
\nv\times\Av = \hat{z} B = \hat{z} m\omega_c = \hat{z} 2m\omega \ .
\end{equation}
In electronic materials flux is created by the external magnetic field $B$, while in ultra-cold atom systems it originates in rotation at angular velocity $\omega$ (the relationship above follows from dynamics in the rotating frame of reference). We will not consider the fluctuations of $\Av$, so that the model will naturally describe neutral particles (although Coulomb forces can be implemented as direct interactions). The imaginary-time action is:
\begin{eqnarray}\label{RotAct}
S & = & \int\dd\tau\dd^2r \Bigl\lbrack \psi_{i\alpha}^\dagger \left( \frac{\partial}{\partial\tau}
    + \frac{(-i\nv-\Av)^2}{2m} - \mu \right) \psi_{i\alpha}^{\phantom{\dagger}}  \nonumber \\
 && + N \Phi^\dagger \hat{\Pi}_0 \Phi
    + \Phi^\dagger \psi_{i\uparrow}^{\phantom{\dagger}} \psi_{i\downarrow}^{\phantom{\dagger}}
    + \Phi \psi_{i\downarrow}^\dagger \psi_{i\uparrow}^\dagger \Bigr\rbrack
    \ .
\end{eqnarray}
The fermionic (Grassmann) fields $\psi_{i\alpha}$ carry spin $\alpha\in\lbrace\uparrow,\downarrow\rbrace$ and a flavor index $i=1\dots N$ which will be used to systematically generate perturbative expansions even when interactions are strong. The physical case of interest is $N=1$. The attractive interactions between particles are decoupled in the particle-particle channel by the Hubbard-Stratonovich field $\Phi$. These interactions lead to processes in which two colliding particles are annihilated and recreated at a different location and different time, so that we can view $\Phi$ as a mediating Cooper pair field. No particular assumptions are made about the nature or range of interactions at this point, so we leave the bosonic kernel $\hat{\Pi}_0$ unspecified until the next section.

It is worthwhile noting that in cold-atom systems the gauge field $\Av$ is rigid and cannot be expelled from a superfluid because it captures the inertial forces in the rotating frame of reference. As a consequence, cold atom superfluids are automatically type-II. In contrast, $\Av$ is dynamical in electronic systems and the type of a superconductor is determined by various parameters not contained in this model, such as disorder.

The action is formally a generalization of the ``two-channel'' model to the Sp($N$) symmetry group. The purpose of this generalization is to systematically organize a perturbation theory with $1/N$ as a small parameter even when interactions are not weak, as will be the case throughout the paper. We first note that the full Cooper pair propagator is proportional to $1/N$. This follows from the Dyson equation whose solution can be represented using Feynman diagrams:
\begin{equation}\label{DysonPhi}
\raisebox{-0.25in}{\includegraphics[height=0.6in]{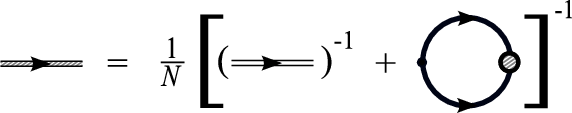}}
\end{equation}
A double line represents $\hat{\Pi}_0^{-1}$, while a single thick line is the full fermion propagator (note that the summation over fermion flavors in the bubble diagram contributes a factor of $N$). The full vertex represented by a shaded dot, and the full fermion propagator are obtained by dressing the bare vertex and propagator with full Cooper pair lines from the left-hand-side, and this generates contributions at higher orders in $1/N$. In general, any physical observable will formally be expressed as an expansion in powers of $1/N$, justifying perturbation theory for any interaction strength in the limit of large $N$. This is a semi-classical expansion because for $N=\infty$ all pairing fluctuations are suppressed and saddle-point approximations become exact.

The uncorrelated Gaussian state to which the pairing interactions are added as perturbations is an integer quantum Hall state of fermions. This state is hard to imagine as a ground state of any interacting system because of its macroscopically Landau-degenerate spectra. Nevertheless, it is a good starting point at sufficiently high temperatures, and its usefulness at zero temperature will become apparent in section \ref{SecVL}.

The bare fermionic states are Landau orbitals with energies $\epsilon_n$ which we will represent in the Landau gauge:
\begin{eqnarray}\label{LandauGauge0}
\Av & = & -B y \xh \\
\epsilon_n & = & \omega_c \left( n + \frac{1}{2} \right) \nonumber
\end{eqnarray}
The quantum numbers are Landau level index $n\in\lbrace 0,1,2\dots \rbrace$ and momentum $k$ in $x$-direction. The bare fermion wavefunctions $\psi_{n,k}$ and the Cooper pair wavefunctions $\Phi_{n,p}$ that will be used to expand the order parameter are:
\begin{eqnarray}\label{LandauGauge}
\psi_{n,k}(\rv) & = & \frac{1}{\sqrt{2^n n!}} \left( \frac{B}{\pi} \right)^{\frac{1}{4}} e^{ikx} \times \\
  && e^{-\frac{B}{2}\left(y+\frac{k}{B}\right)^2} H_n\left(\sqrt{B}y+\frac{k}{\sqrt{B}}\right) \nonumber \\
\Phi_{n,p}(\rv) & = & \frac{1}{\sqrt{2^n n!}} \left( \frac{2B}{\pi} \right)^{\frac{1}{4}} e^{ipx} \times \nonumber \\
  && e^{-B\left(y+\frac{p}{2B}\right)^2} H_n\left(\sqrt{2B}y+\frac{p}{\sqrt{2B}}\right) \ , \nonumber
\end{eqnarray}
($H_n$ are Hermite polynomials). Note that $\Phi_{n,p}$ carries twice the ``charge'' of fermions $\psi_{n,k}$.

The elements of the perturbation theory in Landau representation are summarized in table \ref{Pert}. The dimensionless vertex function in Landau representation is readily derived from ~(\ref{LandauGauge}):
\begin{eqnarray}\label{Vertex1}
&& \Gamma_{m_1,m_2}^n(\xi) = \frac{2^{-(n+m_1+m_2)/2}}{\sqrt{\pi n!m_1!m_2!}}
  \left(\frac{2}{\pi}\right)^{\frac{1}{4}} e^{-\xi^2} \times \\
&&  \times \int_{-\infty}^{\infty} \dd\eta e^{-2\eta^2} H_n(\sqrt{2}\eta) H_{m_1}(\eta+\xi) H_{m_2}(\eta-\xi)
  \nonumber \ .
\end{eqnarray}
Note that the vertex does not depend on the momentum transferred to the Cooper pair, but only on the difference of incoming fermion momenta.

\begin{table}
  \begin{tabular}{c@{\quad}c@{\quad}c}
    \hline \\
    \includegraphics[width=0.6in]{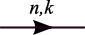} &
      fermion &
      $\displaystyle \frac{1}{-i\omega_m+\epsilon_n}$ \\
    \includegraphics[width=0.6in]{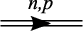} &
      Cooper pair &
      $\displaystyle \hat{\Pi}_0$ \\
    \raisebox{-0.37in}{\includegraphics[width=1.1in]{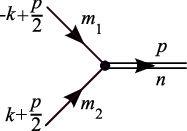}} &
      vertex &
      $\displaystyle \Gamma_{m_1,m_2}^n \left( \frac{k}{\sqrt{B}} \right)$ \\
   \\ \hline
  \end{tabular}
\caption{\label{Pert}Bare elements of the perturbation theory in Landau representation. The vertex function $\Gamma$ is given by ~(\ref{Vertex1}), and $\omega_m$ is Matsubara frequency.}
\end{table}

\subsection{Pairing instability}\label{SecPairing}

Consider bosonic two-particle excitations in the normal phase. Their dynamics is captured by the Cooper pair propagator $\hat{D} = (N\hat{\Pi})^{-1}$. In general, we need to view this propagator as an operator, or a matrix whose rows and columns are indexed by the quantum numbers of Cooper pairs in some representation. We do not know a priori which representation diagonalizes $\hat{\Pi}$, but the Landau representation is a convenient one to work with. Since pairing instability occurs at zero frequency, the easiest way to detect it is by the onset of negative eigenvalues of $\hat{\Pi}$ at zero Matsubara frequency. This follows from the fact that upon integrating out the fermion fields in ~(\ref{RotAct}) the effective Cooper pair action has the quadratic term $\Phi^{\dagger}\hat{\Pi}\Phi$ which needs to be positive for any vector $\Phi$ if the perturbation theory about the normal state ($\Phi=0$) were to be stable.

Before embarking on the calculation of $\hat{\Pi}$ we must note that it is not gauge invariant. Gauge transformations in position representation that leave the action invariant are:
\begin{eqnarray}\label{GaugeTransf}
\Av(\rv) & \to & \Av(\rv) + \nv\lambda(\rv) \\
\psi_{i\alpha}(\rv) & \to & \psi_{i\alpha}(\rv) e^{i\lambda(\rv)} \nonumber \\
\Phi(\rv) & \to & \Phi(\rv) e^{i 2\lambda(\rv)} \nonumber \\
\Pi(\rv_1,\rv_2) & \to & \Pi(\rv_1,\rv_2) e^{i\lbrack 2\lambda(\rv_1) - 2\lambda(\rv_2) \rbrack} \nonumber
\end{eqnarray}
Gauge transformations are manifestly unitary transformations of $\hat{\Pi}$ and do not affect its spectrum of eigenvalues. Therefore, the spectrum which contains all information about bosonic dynamics is gauge invariant and the calculations can safely proceed in any gauge choice. Note that if the gauge field were dynamic one would not be able to apply the standard perturbation theory in powers of charge because the gauge transformations would appear non-perturbative. That problem is absent in our approach.

The Cooper pair kernel $\hat{\Pi}$ can be viewed as a sum of two parts according to ~(\ref{DysonPhi}). The first part is the bare kernel $\hat{\Pi}_0$ which contains information about the microscopic details of the attractive interactions between fermions. The second part is the fermion bubble diagram which can be derived as an expansion in $1/N$:
\begin{eqnarray}\label{Pi1}
&& \Pi'_{n,n'}(p,i\Omega) = \sqrt{B} \sum_{m_1,m_2} \int\frac{\dd k}{2\pi}
  \frac{f(\varepsilon_{m_1}) - f(-\varepsilon_{m_2})}
       {-i\Omega + \varepsilon_{m_1} + \varepsilon_{m_2}} \nonumber \\
&& ~~ \times \Gamma_{m_1,m_2}^{n*}\left(\frac{k}{\sqrt{B}}\right)
  \Gamma_{m_1,m_2}^{n'}\left(\frac{k}{\sqrt{B}}\right)
  + \mathcal{O}\left(\frac{1}{N}\right) \ .
\end{eqnarray}
Note that this expression for $\hat{\Pi}'$ written in the Landau representation is diagonal in $p$ but not in the bosonic Landau level indices $n$. The lowest order term, explicitly written here with the internal Matsubara frequency summed up, contains only the bare fermion propagators and no vertex corrections.

As we will discuss in section \ref{SecRG}, different microscopic details of attractive interactions between fermions can in principle lead to different properties of the system. Without being able to classify all possible cases we still have to choose a particular form of interactions in order to carry out calculations. The simplest choice is to model short-range interactions by a zero-range potential. The Hubbard-Stratonovich decoupling of such attractive interactions produces $\hat{\Pi}_0 = \textrm{const.}$:
\begin{eqnarray}
&& \exp\left\lbrace \frac{V}{N}
  \int\dd\tau \dd^2 r \psi_{i\uparrow}^{\dagger}(\rv) \psi_{i\downarrow}^{\dagger}(\rv)
  \psi_{j\downarrow}^{\phantom{\dagger}}(\rv) \psi_{j\uparrow}^{\phantom{\dagger}}(\rv) \right\rbrace \\
&& ~~\propto \int\mathcal{D}\Phi \exp\Biggl\lbrace -\int\dd\tau \dd^2 r \Biggl\lbrack \frac{N}{V} |\Phi(\rv)|^2
   \nonumber \\
&& ~~~~ + \Phi^{\dagger}(\rv) \psi_{i\uparrow}^{\phantom{\dagger}}(\rv) \psi_{i\downarrow}^{\phantom{\dagger}}(\rv) +
  \Phi(\rv) \psi_{i\downarrow}^{\dagger}(\rv) \psi_{i\uparrow}^{\dagger}(\rv) \Biggr\rbrack \Biggr\rbrace
  \ . \nonumber
\end{eqnarray}
In three dimensions and zero flux density the constant $\hat{\Pi}_0 = V^{-1}$ is related to the scattering length of two body collisions in vacuum: $\hat{\Pi}_0 = m\nu/4\pi$, where $a=-1/\nu$ is the scattering length. In the context of cold atom experiments, $\nu$ is the detuning from the Feshbach resonance. For $\nu=0$ the scattering length diverges and the system acquires universal properties, which is the so called unitarity regime where microscopic details of interactions do not matter and the simple interaction potential above is sufficient for making various quantitative predictions. In two dimensions we must redefine the detuning in order for $\hat{\Pi}_0$ to have proper engineering dimensions: we replace $\nu$ by $\widetilde{\nu} = \nu a_z$, where $a_z$ is a confinement length-scale or ``width'' of the potential well in $z$-direction which effectively reduces the system to two dimensions.

While establishing these connections with cold-atom systems is useful, one should be aware of the conceptual problems in the presence of finite flux density. Strictly speaking, scattering length $a$ is a quantity that describes interactions between extended free particles, but in our case particles are localized in Landau orbitals. Only for $a \ll R_c$, the cyclotron radius at Fermi energy, $a$ retains its physical meaning. Therefore, it is most appropriate to regard $\nu=-1/a$ just as a measure of the strength of interactions, positive values being the BCS limit and negative values being the BEC limit. Interpreting $\nu=0$ as a Feshbach resonance in cold-atom systems and ascribing universal properties to it is problematic, and the discussion in section \ref{SecRG} will shed more light on this.

In fact, the relationship between $\hat{\Pi}_0$ and scattering length is established only after the regularization of the infra-red and ultra-violet behavior of $\hat{\Pi}$. In three dimensions and without the external flux the $\hat{\Pi}$ given by the expression analogous to ~(\ref{Pi1}) is known to be ultra-violet divergent. Once this field-theoretical artifact is removed by regularization, the leftover contribution to $\hat{\Pi}$ is precisely $m\nu/4\pi$ \cite{unitary}. In two dimensions $\hat{\Pi}$ becomes infra-red divergent as well if the bare fermion spectrum is gapless. In our case the spectrum is gapped due to Landau level quantization, but more generally any two-dimensional system should be viewed as a three-dimensional system constrained by a potential well in which the lowest energy band is raised above the bottom of the well. Therefore, infra-red divergence is not a real concern. The ultra-violet behavior is, however, very different from the zero flux case, because the flux modifies the spectrum at arbitrarily high energies. On physical grounds we perform regularization by subtracting from ~(\ref{Pi1}) the same expression $\hat{\Pi}'(0)$ evaluated at zero Matsubara frequency and momentum transfer, as well as zero temperature and chemical potential, but finite flux density. This takes care of any possible ultra-violet divergencies and is compensated by adding $m\widetilde{\nu}/4\pi$ which in the limit of small scattering length is precisely given by $\hat{\Pi}'(0)$. The final expression for the Cooper pair kernel that we shall use is:
\begin{equation}\label{Pi2}
\hat{\Pi} = \frac{m\widetilde{\nu}}{4\pi} + \hat{\Pi}' - \hat{\Pi}'(0) + \mathcal{O}\left(\frac{1}{N}\right) \ .
\end{equation}

\begin{figure}[!]
\includegraphics[height=2.7in]{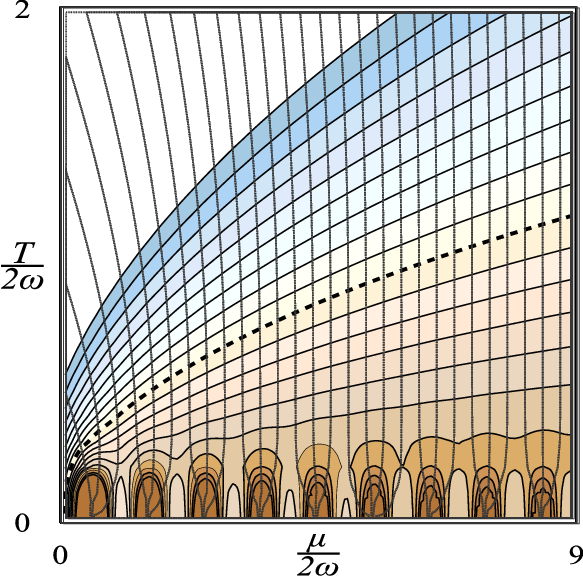}
\caption{\label{Pairing2}(color online) Critical temperature for the onset of Cooper pairing as a function of chemical potential for several values of dimensionless detuning $\widetilde{\nu}=\nu a_z$: the outermost curve is BEC limit $\widetilde{\nu}=-0.9$, the innermost is BCS limit $\widetilde{\nu}=+0.9$, unitarity $\nu=0$ is dashed ($\Delta\widetilde{\nu}=0.1$). The vertically stretching lines are normal state constant density contours, with increment $0.25 B/(2\pi)$.}
\end{figure}

Figure \ref{Pairing2} shows pairing second-order phase transitions in the limit $N\to\infty$ obtained from the onset of negative eigenvalues of ~(\ref{Pi2}). The numerical calculation of ~(\ref{Pi1}) involves the first 500 fermionic Landau levels below the cut-off and the lowest bosonic Landau level, although the phase diagram remains the same if higher bosonic Landau levels $n, n'$ are included. The deeper one goes into the BEC regime, the faster the growth of critical temperature with chemical potential (density). In the BCS regime, however, the paired phase breaks up into dome-shaped islands at low densities, obtained when chemical potential gets close to a fermionic Landau level. Once a paired state becomes a connected region at sufficiently large densities, there are trailing islands of unpaired states sitting between the Landau levels. These details are shown with more clarity in Fig.\ref{PairingPD}. Related phenomena have been considered in literature, but the zero temperature normal states were not anticipated \cite{Zlatko3}.

Integer quantum Hall states are obtained by putting the chemical potential $\mu$ between Landau levels. Pairing interactions compete with the cyclotron gap $\omega_c=2\omega$ and must be strong enough in comparison to the smallest single-particle gap $\textrm{min}\lbrace |\mu-\omega_c(n+\frac{1}{2})| , n\in\mathcal{Z}\rbrace$ in order to give rise to Cooper pairing. This means that if pairing is not too strong, insulating states can exist at zero temperature with fully populated Landau levels below the chemical potential. The analysis based on pairing instability in the perturbation theory is, however, not complete. In the next two sections we discuss what has been missed.

\subsection{Limitations of the perturbation theory}\label{SecLim}

The perturbation theory seemingly discovers Cooper pairing instability for any finite $N$ when the chemical potential is brought sufficiently close to a Landau level. To cure this instability one usually redefines the ground-state by introducing an order parameter and then rebuilds the perturbation theory about the new ground-state. The order parameter which describes a superfluid with broken U(1) symmetry and an Abrikosov lattice of vortices can be written as a superposition of Landau level wavefunctions in ~(\ref{LandauGauge}):
\begin{equation}
\Phi_0(\rv) = \sum_n \int\frac{\dd p}{2\pi} \phi_{n,p} \Phi_{n,p}(\rv) \ .
\end{equation}
Hence, the order parameter is specified by multiple amplitudes $\phi_{n,p}$. If we substitute $\Phi(\rv) = \Phi_0(\rv) + \delta\Phi(\rv)$ in ~(\ref{RotAct}) and integrate out the fluctuating fields $\psi$ and $\delta\Phi$ we can obtain the perturbative $1/N$ expansion for free energy density of the ansatz state characterized by $\Phi_0$:
\begin{equation}\label{FreeEnergy}
\frac{\mathcal{F}(\Phi_0)}{N} = \frac{\mathcal{F}_0}{N} + \hat{\Pi}_{ij}\Phi_0^{i*}\Phi_0^j
   + \hat{U}_{ijkl}\Phi_0^{i*}\Phi_0^{j*}\Phi_0^k\Phi_0^l + \mathcal{O}(\Phi^6) \ .
\end{equation}
The quantum numbers $(n,p)$ are represented by indices $i,j\dots$ for brevity, and all couplings $\hat{\Pi}, \hat{U}\dots$ are functions of $\mu, T, \omega\dots$ which have tensor structure and contributions to all powers of $1/N$. Note that the expansion is analytic at $\Phi_0=0$ because the unperturbed state at $\Phi_0=0$ is a fully gapped incompressible quantum Hall state of unpaired fermions. In the normal state $\mathcal{F}$ is minimized at $\Phi_0=0$, while in the superfluid the minimum is at $\Phi_0 \neq 0$.

At the lowest order or the mean-field level we found both superfluid and normal states in the previous section. If we include higher order corrections the order parameter in a superfluid state will be reduced and the normal areas in the phase diagram will grow. However, superfluid states will survive at all orders of perturbation theory for any finite $N$ if the chemical potential is sufficiently close to a Landau level. This is ultimately a consequence of the cyclotron gap in the fermion spectrum which prevents the occurrence of non-analytic infra-red features and yields finite corrections to all couplings in ~(\ref{FreeEnergy}) at all orders of $1/N$. Alternatively, any point in the phase diagram which is a superfluid in the mean-field approximation will remain superfluid for sufficiently large $N$ despite fluctuations.

Throughout this paper we exploit the computational and analytical convenience of the physical picture embodied in  ~(\ref{FreeEnergy}). However, we must be aware of its limitations. This picture naively suggests two kinds of broken symmetries in the superfluid phase, the ``off-diagonal'' long-range order and space-group symmetry breaking due to the vortex lattice. Even at finite temperatures it suggests the existence of these broken symmetries. On the other hand, it is well known that fluctuations in two-dimensional systems can restore continuous symmetries in the equilibrium states. In particular, the superfluid phase at zero temperature has a long-range ordered vortex lattice, but shear fluctuations of vortex positions reduce the ``off-diagonal'' order to algebraic correlations and hence restore the U(1) symmetry \cite{Moore, Zlatko2, MacDonald}. At finite temperatures even the vortex lattice must be melted according to Mermin-Wagner theorem, but algebraic correlations of vortex positions are allowed and Kosterlitz-Thouless transitions can separate the quasi long-range ordered vortex ``lattice'' phases from disordered phases with only short-range correlations.

The naive approach behind the Eq.\ref{FreeEnergy} assumes that the only two kinds of states one needs to worry about are the normal state with short-range correlations of the order parameter, and the superfluid with long-range correlations. It does not explore the free energy density $\mathcal{F}'$ of states with \emph{algebraic} correlations of the order parameter(s). It turns out that in the circumstances mentioned above $\mathcal{F}'<\mathcal{F}(\Phi_0)$ for any attempted order parameter $\Phi_0$ inside the naively predicted superfluid regions. Another possibility is that $\mathcal{F}'<\mathcal{F}(0)$ inside some parts of the naively predicted normal states.

Calculating $\mathcal{F}'$ is highly non-trivial in the perturbation theory. Therefore, we are better off learning what we can from $\mathcal{F}(\Phi_0)$ in ~(\ref{FreeEnergy}). The difference $\Delta = \mathcal{F}(\Phi_0) - \mathcal{F}'$ defines an energy scale associated with the error made by ~(\ref{FreeEnergy}). We will be able to make conclusions only about phenomena characterized by energies larger than $\Delta$, or length-scales smaller than $\delta \propto \Delta^{-1}$. This will be sufficient for establishing the existence of zero temperature vortex liquid phases in the next section. On general grounds we can expect that the error $\Delta$ is small because even in the absence of a true long range order the scale given by $\Phi_0$ plays a role in various properties of the system.


\subsection{Vortex liquid}\label{SecVL}

The quantum Hall insulator of unpaired fermions is a very peculiar unperturbed state for building the perturbation theory. A very important property of the \emph{exact} Cooper pair propagator $\hat{D}=(N\hat{\Pi})^{-1}$ in this state is that its spectrum is macroscopically degenerate and corresponds to bosonic Landau level degeneracy. A simple argument is that in the absence of any emerging parameters that could characterize the equilibrium state the propagator $\hat{D}$ obtains its matrix structure from the only single-particle operator available in the theory ~(\ref{RotAct}), the canonical momentum $\Pv=-i\nv-\Av$. This operator has a degenerate spectrum that gives rise to Landau level degeneracy, hence the operator $\hat{D}(\Pv)$ must also have a degenerate spectrum, assuming of course that the functional dependence on $\Pv$ is analytic. A detailed demonstration of this fact follows from perturbation theory to all orders, and agrees with a general expectation that the perturbation theory describes a state smoothly connected to the state of non-interacting particles in the absence of instabilities. It is important to note here that Cooper pairs are ``charged'' with respect to the gauge field; ``charge-neutral'' particle-hole excitations are exempt from the Landau-level quantization.

\begin{figure}[!]
\includegraphics[height=0.8in]{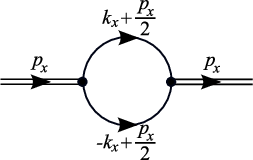}
\hspace{0.2in}
\raisebox{0.15in}{\includegraphics[height=0.7in]{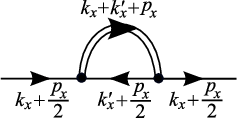}}
\caption{\label{SEexample} Feynman diagram labeling examples.}
\end{figure}

The perturbative argument goes like this. We can establish that ~(\ref{Pi1}) and hence ~(\ref{Pi2}) are independent of $p$ to all orders of $1/N$. First, the written lowest order term in the $1/N$ expansion involves fermion energies that do not depend on $\pm k + p/2$, and vertices ~(\ref{Vertex1}) that do not depend on $p$. This term defines the bare boson propagator $D_0 \propto N^{-1}$, using which we generate higher order corrections of ~(\ref{Pi1}). Let us apply the following labeling rules in all Feynman diagrams that contribute to ~(\ref{Pi2}): each fermion propagator shall carry momentum $k+p/2$ in the arrow direction, while each boson propagator shall carry $q+p$ (see Fig.\ref{SEexample}). The transfer of $p$ is automatically conserved. Each added vertex takes momenta $k_1+p/2$ and $k_2+p/2$ at its fermionic terminals, but there is no dependence on $p$ since only the difference of momenta at the fermion terminals matters in ~(\ref{Vertex1}). The added bare boson lines take $p$ contributions, but according to ~(\ref{Pi1}) they do not depend on the transferred momentum. Therefore, order by order, no dependence on $p$ is introduced in any Feynman diagram.

Changing the nature of interactions does not modify this conclusion because $\hat{\Pi}_0$ must be gauge invariant in the presence of the external uniform flux density. Even interactions in the particle-hole channel cannot lift the degeneracy of bosonic modes (an equivalent ``degeneracy theorem'' can be proved to all perturbative orders in a Landau-Ginzburg theory of interacting particles in magnetic field). Therefore, the gapped two-particle bosonic modes are macroscopically degenerate in the trivial insulating state regardless of the type of interactions between particles. Naturally, it is hard to believe that any quantum fluctuations would not lift this degeneracy. At no surprise we shall demonstrate that the degeneracy must be lifted at least close to the superfluid transition, but the main point for now is that it takes spontaneous reorganization of the ground-state to lift the degeneracy (which in some sense is a trivial but very important observation). Even without knowing whether the obtained normal $T=0$ state has some conventional order or topological order, it is certain that it cannot be smoothly connected with the ordinary insulating state.

First, we need to understand the properties of the zero temperature superfluid phase. Let us start from a quantum Hall insulator at zero temperature. When the energy of the lowest degenerate bosonic states in the insulator becomes negative in response to the chemical potential approaching a Landau level, then Cooper pairs condense into those states. The resulting order that minimizes free energy density is automatically an Abrikosov vortex lattice residing initially just in the lowest bosonic Landau level. The free energy cost of rearranging vortices without changing the vortex density is exceptionally small near the phase transition. By assumption we are dealing with type-II superfluids so that the pairing phase transition is second order. But, even more striking is the effect of bosonic degeneracy in the trivial insulating phase.

Consider the free energy density as a function of order parameter in ~(\ref{FreeEnergy}). We argued that the quadratic coupling $\hat{\Pi}$ is macroscopically degenerate. Therefore, different vortex lattice arrangements, specified by different order parameters $\Phi_0$, will differ in free energy only as $\mathcal{O}(\Phi_0^4)$ close to the phase transition. We conclude that the potential energy density of a vortex lattice depends on the order parameter near the transition as:
\begin{equation}\label{Epot}
E_{\textrm{pot}} \sim \Phi_0^4 \ .
\end{equation}

Quantum fluctuations of the order parameter dynamically distort the vortex lattice. Their microscopic origin are the short-range repulsive interactions between Cooper pairs which are caused at least by the Pauli exclusion of their fermionic constituents. Interactions between bare fermions also affect the quantum fluctuations in a manner that can be very sensitive to microscopic details of the interaction potential. In fact, all possible two-body and multi-particle interactions are relevant at the Gaussian fixed point in the renormalization group sense, regardless of their spatial dependence (see next section). Having infinitely many relevant directions in the parameter space suggests that there could be many interacting stable fixed points that describe numerous possible correlated phases. Systematically analyzing all possibilities seems hopeless. Therefore, we shall not attempt to identify the properties of strongly correlated normal phases that can be obtained after the superfluid is destroyed by quantum fluctuations. Instead, we can at least prove that \emph{some} strongly correlated normal phase \emph{must} inherit the superfluid, rather than the trivial unpaired quantum Hall insulator. This is already one statement about the renormalization group flow of parameters near the superfluid quantum phase transition.

In order to summarize the effects of quantum fluctuations we can view them as quantum motion of vortices. In this picture, vortex lattice melting destroys the superfluid. It will suffice to estimate the effective vortex mass $M_{\textrm{v}}$, specifically its dependence on the order parameter $\Phi_0$. We make the estimate using the Heissenberg uncertainty relation. Vortices are localized in a vortex lattice at the scales of magnetic length $l_{\textrm{m}} = 1/\sqrt{m\omega_c}$ which measures the average distance between vortices and corresponds to a momentum scale $p_{\textrm{m}} \sim \sqrt{m\omega_c}$. If vortices were allowed to move freely they could convert the potential energy of localization in the lattice to kinetic energy of the order of $p_{\textrm{m}}^2/2M_{\textrm{v}}$ per vortex. Since vortex density is topologically fixed by the constant rotation rate or magnetic field, we find:
\begin{equation}\label{Ekin0}
E_{\textrm{kin}} \sim M_{\textrm{v}}^{-1}(\Phi_0) \ .
\end{equation}

In typical fermionic superfluids one is usually mostly concerned with the quasiparticle contribution to vortex mass. In $s$-wave superfluids vortex cores can localize quasiparticles, so that $M_{\textrm{v}}$ is at least of the order of the total mass of trapped core fermions. An upper bound for $M_{\textrm{v}}$ can be found by viewing a vortex core as a potential well of depth $|\Phi_0|$ in the units of energy, and radius comparable to the Cooper pair coherence length $\xi\sim|\Phi_0|^{-1}$. Assuming roughly a constant density of core states appropriate for two-dimensional non-relativistic quasiparticles, the total number of core states is proportional to $|\Phi_0|$, so that $M_{\textrm{v}} \lesssim |\Phi_0|$. This estimate is best suited for the strong BCS limit \cite{Kopnin}. As the strength of interactions is increased, the coherence length decreases, quasiparticles are gradually being expelled from vortex cores and $M_{\textrm{v}}$ is being reduced. There is also a contribution of extended quasiparticles to vortex mass. In $s$-wave superfluids this contribution is thermally activated, but in $d$-wave superfluid it gives rise to a vortex mass of the order of a fermion mass \cite{DWVorDyn}.

Close to the superfluid transition it turns out that the ``hydrodynamic'' vortex mass due to the phase fluctuations of the order parameter dominates in $M_{\textrm{v}}$ (for any finite $N$). This contribution can be generally estimated as $M_{\textrm{v}} \sim \epsilon_{\textrm{v}}/c^2$, where $\epsilon_{\textrm{v}}$ is energy of a static vortex and $c$ is the speed of sound, or the superfluid critical velocity \cite{DWVorDyn, Duan1}. For a neutral superfluid with stiffness $\rho_{\textrm{s}}$ in the units of energy, $c \sim \sqrt{\rho_{\textrm{s}}/m}$ while $\epsilon_{\textrm{v}} \propto \rho_{\textrm{s}} \log(l_{\textrm{m}}/\xi)$ diverges logarithmically with the separation $l_{\textrm{m}}$ between vortices. The logarithmic behavior comes from the contribution of the circulating supercurrent phase gradients that extend beyond the vortex core of radius $\xi$ freely up to the distance to the neighboring vortices (we assumed that the London penetration length is larger than the magnetic length $l_{\textrm{m}}$). Using $\rho_s \sim |\Phi_0|^2$ we find that the hydrodynamic contribution to the vortex mass behaves as $M_{\textrm{v}} \sim \log|\Phi_0|$ near the phase transition for a fixed flux density. In charged superfluids the logarithmic factor is screened-out \cite{DWVorDyn, Duan2}, so that the vortex mass approaches a constant as $|\Phi_0|\to 0$. Therefore, the vortex kinetic energy density is:
\begin{equation}\label{Ekin}
E_{\textrm{kin}} \sim \left\lbrace
\begin{array}{l@{\quad,\quad}l}
  \left\vert \log|\Phi_0| \right\vert^{-1} & \textrm{neutral superfluid} \\
  \textrm{const.} & \textrm{superconductor}
\end{array}
\right\rbrace \ .
\end{equation}

Regardless of the concrete microscopic details and factors that enter ~(\ref{Epot}) and ~(\ref{Ekin}), close enough to the pairing transition the kinetic energy wins and vortex lattice must melt. The quantum melting occurs at a finite $|\Phi_0|$ below which $E_{\textrm{kin}} > E_{\textrm{pot}}$, before the depairing transition is reached. Both the kinetic and potential energy of vortices can be calculated in the perturbation theory, and the melting transition is expected to be first order according to the Landau-Ginzburg theory of phase transitions. The resulting normal state can be regarded a strongly correlated vortex liquid in which Cooper pairs remain the most important low energy degrees of freedom. The energy released into the motion of vortices in the vortex liquid is not sufficient to overcome the pairing gap $|\Phi_0|$ in the fermion spectrum.

By comparing the spectrum of bosonic excitations in this vortex liquid and the trivial quantum Hall insulator one easily concludes that the two normal phases cannot be smoothly connected. The bosonic modes in the superfluid state close to the pairing transition are nearly degenerate, so that many different vortex arrangements have very similar free energy densities. In the trivial quantum Hall insulator all of these modes become gapped and collapse to the same bosonic Landau level, leaving the occupied fermionic Landau levels to define the ground-state. On the other hand, vortex liquid is obtained by mixing the nearly-degenerate bosonic modes inside the superfluid, while they still have negative energies that encourage condensation. The mixing perturbation produces a new ground-state (with lower energy than the superfluid) as a quantum superposition of many different vortex arrangements and pushes away from it the other collective modes toward higher energies. The resulting low energy spectrum is qualitatively different than that of the trivial quantum Hall insulator because the ground-state is still a collective state of correlated bosons at some negative energy where condensation is prevented only due to strong quantum fluctuations.

We cannot predict the properties of the vortex liquid, and they are not universal. However, the perturbation theory requires that a vortex liquid be adjacent to any type-II superfluid phase of a two-dimensional quantum fermionic system at zero temperature with explicitly broken time-reversal symmetry. How should this be interpreted in the light of the fact that the naive perturbation theory does not correctly describe the lowest energy scales? Even though quantum fluctuations destroy the superfluid long-range order, they leave behind the broken symmetries due to the vortex lattice. The vortex lattice potential energy ~(\ref{Epot}) is shaped by quasiparticle and Cooper pair mediated forces between vortices, which in turn are well defined at sufficiently short length-scales to be captured by the naive perturbation theory. Specifically, close to the pairing transition when the error length-scale $\delta$ becomes large in comparison with the separation $l_{\textrm{m}}$ between vortices the naive estimate ~(\ref{Epot}) can be trusted. Similarly, vortex mass is always dominated by the ``high energy'' degrees of freedom below the ultra-violet cut-off associated with the finite vortex core size, or the superconducting ``gap''. Vortex is a local deformation of superfluid phases so that its dynamics can be reliably captured by the naive perturbation theory. Therefore, we can trust the estimate ~(\ref{Ekin}) as well and establish the existence of a vortex liquid at zero temperature.

\subsection{Renormalization group analysis}\label{SecRG}

The purpose of the following discussion is to elucidate the issues behind attempts to apply the simple theory ~(\ref{RotAct}) to the analysis of realistic systems. A usual theoretical strategy is to justify a tractable field theory in the vicinity of second-order phase transitions where most operators allowed by symmetries do not significantly affect the macroscopically observable properties of the system. The fact that only a few operators are important to keep in calculations is what makes a field theory quantitatively useful. The model ~(\ref{RotAct}) does not have this useful property.

We demonstrate using renormalization group (RG) that all arbitrary-range interactions between fermions are perturbations to ~(\ref{RotAct}) that can shape phases and induce transitions. Physically, this comes from the fact that the bare fermion spectrum in ~(\ref{RotAct}) consists of dispersionless macroscopically degenerate Landau levels with energies $\epsilon_n = n \omega_c - \mu'$ (here we measure energy with respect to the redefined chemical potential $\mu'=\mu-\omega_c/2$). The lack of dispersion reduces dimensionality in RG and makes even arbitrarily weak perturbations extremely potent in lifting the degeneracy. We set up RG by generalizing ~(\ref{RotAct}) to $d$ dimensions in the Landau level basis using the Landau gauge, and include additional allowed terms:
\begin{widetext}
\begin{eqnarray}\label{RotAct2}
S & = & \int\dd\tau\dd^{d-2}r_{\perp} \Biggl\lbrace \sum_n
        \int\frac{\dd k_x}{2\pi} \psi_{n,k_x}^{\dagger}
          \left( \frac{\partial}{\partial\tau} + n\omega_c - \frac{\nv_{\perp}^2}{2m} - \mu' \right)
        \psi_{n,k_x}^{\phantom{dagger}}
    +   N \sum_{n_1 n_2} \int\frac{\dd p_x}{2\pi}
        \Phi_{n_1,p_x}^{\dagger} \hat{\Pi}_{n_1,n_2}^{(0)} \Phi_{n_2,p_x}^{\phantom{\dagger}}
        \nonumber \\
  & + & g \sum_{n m_1 m_2} \int\frac{\dd k_x}{2\pi} \frac{\dd p_x}{2\pi}
        \Gamma^n_{m_1 m_2} \left( \frac{k_x}{\sqrt{B}} \right) \left\lbrack \Phi_{n,p_x}^{\dagger}
          \psi_{m_1,k_x+\frac{p_x}{2}}^{\phantom{\dagger}}
          \psi_{m_2,-k_x+\frac{p_x}{2}}^{\phantom{\dagger}}
        + \Tr{h.c.} \right\rbrack \nonumber \\
  & + & u_2 \sum_{m_1 \dots m_4} \int\frac{\dd k_{x1}}{2\pi} \frac{\dd k_{x2}}{2\pi} \frac{\dd q_x}{2\pi}
        \Gamma_{m_1 \dots m_4}' \left( k_{x1}, k_{x2}, q_x \right)
        \psi_{m_1,k_{x1}}^{\dagger} \psi_{m_2,k_{x2}}^{\dagger}
        \psi_{m_3,k_{x2}+q_x}^{\phantom{\dagger}} \psi_{m_4,k_{x1}-q_x}^{\phantom{\dagger}}
        \Biggr\rbrace + \cdots \ .
\end{eqnarray}
\end{widetext}
We suppress spin $\alpha$ and flavor $i$ indices for brevity. Since the bare fermion states are localized in the plane perpendicular to the axis of rotation, it is appropriate to not rescale $k_x$ in RG. At zero temperature and density, fermion self-energy vanishes so that all renormalization comes from the boson field $\Phi$. The RG flow equations can be calculated exactly to all orders of perturbation theory when $u_2$ and the remaining omitted couplings are zero, since then the renormalization of $\Phi$ involves summation of a geometric series of bare fermion bubble diagrams. Following the procedure in Ref.\cite{unitary}, we find that under rescaling
\begin{eqnarray}
r_{\perp}'=e^{-l}r_{\perp} ~~ && ~~ \tau' = e^{-2l}\tau \\[0.1in]
\psi'=e^{(d/2-1)l}\psi ~~ && ~~ \Phi'=e^{(d/2-1)l}\Phi \nonumber
\end{eqnarray}
the exact flow equations are:
\begin{eqnarray}
\frac{\dd\mu'}{\dd l} = 2\mu' ~~ && ~~ \frac{\dd\omega_c}{\dd l} = 2\omega_c \\
\frac{\dd\nu}{\dd l} = \left( 2 - a g^2 \right) \nu ~~ && ~~
  \frac{\dd g}{\dd l} = \left( 3 - \frac{d}{2} \right) g - b N g^3 \ , \nonumber
\end{eqnarray}
where $a$ and $b$ are cut-off dependent constants whose values are not important for the present discussion. Appart from the Gaussian fixed point at $\mu'=\omega_c=\nu=g=0$, we can identify an interacting fixed point at $\mu'=\omega_c=\nu=0$, $g^* = \sqrt{(3-d/2)/(bN)}$ which defines the unitarity limit in any dimension $d$, and shows how perturbation theory can be justified for large $N$. In addition to $\mu'$, $\omega_c$ and $\nu$, various couplings $u_n$, which may include non-local short-range potentials and hence multi-particle collision terms, can be relevant at this fixed point. For $n$-particle scattering $u_n$ we find sufficient indication at the tree-level:
\begin{equation}
\frac{\dd u_n}{\dd l} = \bigl\lbrack d + (2-d) n \bigr\rbrack u_n + \mathcal{O}(u^2)
\end{equation}
irrespective of the spatial dependence of interaction potentials in the plane perpendicular to rotation axis, since coordinates do not rescale in this plane. Therefore, all $u_n$ are relevant in $d \le 2$, while in $d>2$ they are relevant for $n<d/(d-2)$.

The same conclusions have been reached in the past using $\epsilon=6-d$ expansions in the context of classical Landau-Ginzburg theories \cite{Nelson}. A formal advantage of the present approach is that the conclusions are justified in any number of dimensions as long as $N$ is large, but also for not so large $N$ as long as the $u_n$ couplings are small and the density of particles is close to zero.

The existence of infinitely many relevant directions in the parameter space near the Gaussian and unitarity fixed points implies the possibility that there could be many stable interacting fixed points that specify the properties of many different phases. The microscopic details of interactions between particles can then very sensitively determine the concrete properties of a concrete system at low temperatures. We expect that the properties of the vortex liquid in particular are not universal. The vortex lattice melting transition lines are also not universal, being first order. The superfluid phase is generally amenable to phenomenological descriptions regardless of the microscopic details, but we can expect that the structure of vortex cores is system dependent.

\subsection{Conclusions}\label{SecCon}

We analyzed a simple model of neutral fermionic particles in two dimensions with attractive interactions and explicitly broken time-reversal symmetry. The model is routinely realized in ultra-cold atom experiments with fermionic atoms near a broad two-body Feshbach resonance, but it also captures some essential properties of much more complicated electronic systems. We assumed that the low-temperature superfluid is type-II so that it hosts an Abrikosov lattice of vortices at a fixed flux density. Using a semi-classical perturbative expansion we found that a second order phase transition separates the superfluid from a normal phase of thermally excited unpaired fermions in the quantum Hall regime. We calculated the critical temperature as a function of interaction strength and density and showed that in some circumstances for weak interactions this phase transition can occur even at zero temperature.

The unusual properties of the quantum Hall insulator at zero temperature were then exploited to argue that the second order superfluid-insulator transition obtained in the naive perturbation theory must be preempted by a first order transition due to quantum vortex lattice melting. This conclusion followed from the analysis of quantum fluctuations related to vortex dynamics. We compared the potential energy of a vortex lattice with the kinetic energy of a vortex liquid and showed that the latter always wins on the superfluid side close enough to the pairing transition. The vortex liquid is a strongly correlated normal phase which cannot be smoothly connected with the quantum Hall state of unpaired fermions.

Finally, we argued that the properties of a vortex liquid in two dimensions at a finite flux density are not universal. Different materials can exhibit different normal states resulting from vortex lattice quantum melting, including states with topological order such as quantum Hall states of Cooper pairs, or states with broken symmetries involving density waves for example. At zero temperature we expect that such states are insulators. At finite temperatures, below a crossover temperature, the normal states will be at least influenced by the proximate non-trivial zero-temperature phases, but we cannot rule out the existence of genuine finite temperature vortex liquids (depending on the broken symmetries of the zero temperature insulator) which are separated from the disordered normal phase by a Kosterlitz-Thouless transition.

Ultra-cold atom experiments have the potential to directly explore the phenomena discussed in this paper. The target regime would be a combination of fast rotation and low particle density where a few lowest Landau levels would be occupied in the quantum Hall state at low temperatures. It is in this limit and on the BCS side of the Feshbach resonance where multiple isolated superfluid phases could be found separated by normal states as the density of particles or rotation rate is varied. In order to faithfully represent our model the experiment would need to achieve this limit with a trap whose harmonic frequency is not much higher than the rotation rate. The phenomena predicted here might survive even in more confining traps, but the structure of vortex lattices and hence the phase boundaries would be affected by the trap in a manner that we cannot predict with a uniform model. Observing Cooper pairs at low temperatures in the normal states would be a first weak indicator of the proximity to a vortex liquid.

\subsection{Acknowledgements}

I am indebted to Olexei Motrunich, Gil Refael, Anton Burkov, Arun Paramekanti, Carlos Bolech and Satyan Bhongale for very helpful discussions. Numerical calculations were performed on Rice University supercomputers. This research was supported by W. M. Keck Program in Quantum Materials.

\end{document}